# Applying configurational complexity to the 2D Ruddlesden-Popper crystal structure

Wenrui Zhang,[1] Alessandro R. Mazza,[1] Elizabeth Skoropata,[1] Debangshu Mukherjee,[2] Brianna L. Musico,[3] Jie Zhang,[1] Veerle Keppens,[3] Lihua Zhang,[4] Kim Kisslinger,[4] Eli Stavitski,[5] Mathew Brahlek,[1] John W. Freeland,[6] Ping Lu,[7] Thomas Z. Ward[1,*]

1. Materials Science and Technology Division, Oak Ridge National Laboratory, Oak Ridge, Tennessee 37831, United States
2. Center for Nanophase Materials Sciences, Oak Ridge National Laboratory, Oak Ridge, Tennessee 37831, United States
3. Department of Material Science and Engineering, University of Tennessee, Knoxville, Tennessee 37996, United States
4. Center for Functional Nanomaterials, Brookhaven National Laboratory, Upton, New York 11973, United States
5. National Synchrotron Light Source II, Brookhaven National Laboratory, Upton, New York 11973, United States
6. Advanced Photon Source, Argonne National Laboratory, Argonne, Illinois 60439, USA
7. Sandia National Laboratory, Albuquerque, New Mexico, 87185, United States

**Abstract**

The 2D layered Ruddlesden-Popper crystal structure can host a broad range of functionally important behaviors. Here we establish extraordinary configurational disorder in a two-dimensional layered Ruddlesden-Popper (RP) structure using entropy stabilization assisted synthesis. A protype $A_2CuO_4$ RP cuprate oxide with five components (La, Pr, Nd, Sm, Eu) on the $A$-site sublattice is designed and fabricated into epitaxial single crystal films using pulsed laser deposition. By comparing $(La_{0.2}Pr_{0.2}Nd_{0.2}Sm_{0.2}Eu_{0.2})_2CuO_4$ crystals grown under identical conditions on several different substrates, it is found that heteroepitaxial strain plays an important role in crystal phase formation during synthesis. When grown on a near lattice matched substrate, the high entropy oxide film features a $T'$-type RP structure with uniform $A$-site cation mixing and square-planar $CuO_4$ units; however, growing under strong compressive strain results in a single crystal non-RP cubic phase consistent with a $CuX_2O_4$ spinel structure. These observations are made with a range of combined characterizations using X-ray diffraction, atomic-resolution scanning transmission electron microscopy, energy-dispersive X-ray spectroscopy, and X-ray absorption spectroscopy measurements. The ability to manipulate configurational complexity and move between 2D layered RP and 3D cubic crystal structures in this class of cuprate materials opens many opportunities for new design strategies related to functionalities, such as magnetoresistance, unconventional superconductivity, ferroelectricity, catalysis, and ion transport.

# 1. Introduction

Ruddlesden–Popper (RP) oxides represent a versatile structural platform and offer various attractive functionalities, including unconventional superconductivity,[1-3] colossal magnetoresistance,[4] tunable dielectrics,[5] efficient electrocatalysis,[6,7] and electrochemical energy conversion.[8-10] The RP crystal motifs allow an extraordinary level of structural flexibility that can be exploited to create and manipulate physical properties sensitive to dimensionality, orbital hybridization, and charge coordination. As described by the format of $A_{n+1}B_nO_{n+1}$,[11] it is possible to modify the number of the intermediate stacking unit $n$ via thermodynamically-governed solid-state synthesis[12,13] or controlled layer-by-layer growth.[14,15] In RP nickelates and cuprates, oxygen stoichiometry engineering is an attractive method to obtain high ionic conductivity for solid oxide fuel cells.[17,18] In most of these materials, active work is underway in which chemical substitution on the perovskite layer is used to control or improve functional responses.

The incorporation of dopants introduce extra configurational disorder to the pristine lattice, by modifying the spin/charge/orbital/lattice degeneracies at the local scale.[19-21] Controlling configurational disorder beyond the conventional substitutional regime could enable dramatic flexibility of material design. However, the difficulty of stabilizing single crystal materials hosting more than a single element on the *A*-site sublattice beyond some small percentage is challenging—crystallization processes during synthesis tend to drive phase segregation as chemical composition complexity increases. Recently, the strategy of entropy assisted synthesis has been applied to transition metal oxides, in order to stabilize multiple elements in a single cation sublattice, forming so-called high-entropy oxides (HEOs).[22-27] This strategy has been effective in stabilizing a number of HEO ceramics with cubic structures, including rock-salt, fluorite, perovskite, and spinel.

The ability to combine high levels of configurational complexity with the inherent benefits of the RP phase is expected to have a profound impact on the functionalities described above. Nonuniformity in the perovskite layer allows fine control over functional behavior through changes in phonon dynamics, spin ordering, charge localization, etc.[28] The cuprate RP class is an excellent example of this, as disorder from the perovskite layer is well known to influence the electron-phonon and electron-spin couplings that drive nonconventional superconductivity.[29,30] Further, combining the exceptional ion conductivity of RP phases with superionic conductivity observed in highly configurationally disordered materials[17,31] could provide new levels of efficiency in solid oxide fuel cell and battery applications. In this study, we demonstrate epitaxial stabilization of a protype $A_2CuO_4$ RP cuprate oxide with five equiatomic $A$-site components (La, Pr, Nd, Sm, Eu) over a range of substrates using pulsed laser deposition. Heteroepitaxial strain is found to play a significant role in stabilizing crystal phase of epitaxial films synthesized under identical conditions. Films grown on near lattice-matched substrates stabilize in the n = 1 RP structure, while films grown under strong compressive strain possess an unexpected non-RP cubic phase. The ability to incorporate elemental diversity into the perovskite layer will allow much finer tuning of order parameters, which not only provide new functionalities in the perovskite layer, but also facilitate modification of the behavior of the 2D cation-oxygen interlayer residing between perovskite layers.

## 2. Results and Discussion

### 2.1 Epitaxial growth of 5ACuO RP-phase films

Epitaxial $(La_{0.2}Pr_{0.2}Nd_{0.2}Sm_{0.2}Eu_{0.2})_2CuO_4$ (5ACuO) films are grown on a range of substrates using pulsed laser deposition (PLD). The five lanthanides on the $A$-site sublattice each have a preferred oxidation state of $3^+$ and similar cation radius ($r_A$). This leads to the selected $A$-site cation combination, giving an average $r_A$ of 1.108 Å and an average variance ($\sigma^2$) of 0.0014. Despite the chemical complexity, excellent single-crystal Ruddlesden-Popper (RP) epitaxial films can be stabilized on near-lattice-matched substrates. Fig. 1a shows $\theta$-$2\theta$ XRD scans of a 20 nm thick 5ACuO epitaxial film grown on a $SrTiO_3$ (STO) (001) substrate. The film is c-axis oriented along the out-of-plane direction, as demonstrated by the presence of only (00$l$) diffraction peaks. Excellent film uniformity leads to well-defined Laue oscillations (Fig. 1b) and periodic interference oscillations in the X-ray reflectivity (XRR) data (Fig. 1c). Asymmetric XRD measurements determine that the 5ACuO film has a nearly identical orthorhombic structure to the well-studied singly $A$-site populated $Nd_2CuO_4$ and $La_2CuO_4$ materials. As seen from the φ-scan measurement, the 5ACuO film grown on the STO substrate presents a fourfold symmetry for its asymmetric <10$\bar{1}$1> diffraction peak (Fig. 1d). Analysis of reciprocal space maps near two orthogonal <10$\bar{1}$1> directions determine the lattice parameters of 5ACuO films (Fig. 1e and Fig. S1, Supporting Information). The lattice mismatch between STO and 5ACuO induce biaxial strain on the film. Increasing the film thickness to 50 nm leads to relaxation in the film, but single-phase, uniform orientation is maintained (Fig. S2, Supporting Information). The lattice parameters of the relaxed film are calculated to be $a$ = 3.87 Å, $b$ = 3.95 Å, and $c$ = 12.21 Å, which are very close to their bulk values found in the ceramic target (Fig. S3, Supporting Information). Atomic force microscopy measurements demonstrate a flat and atomically smooth step-terrace morphology with a surface roughness (Ra) less than 0.5 nm (Fig. 1f).

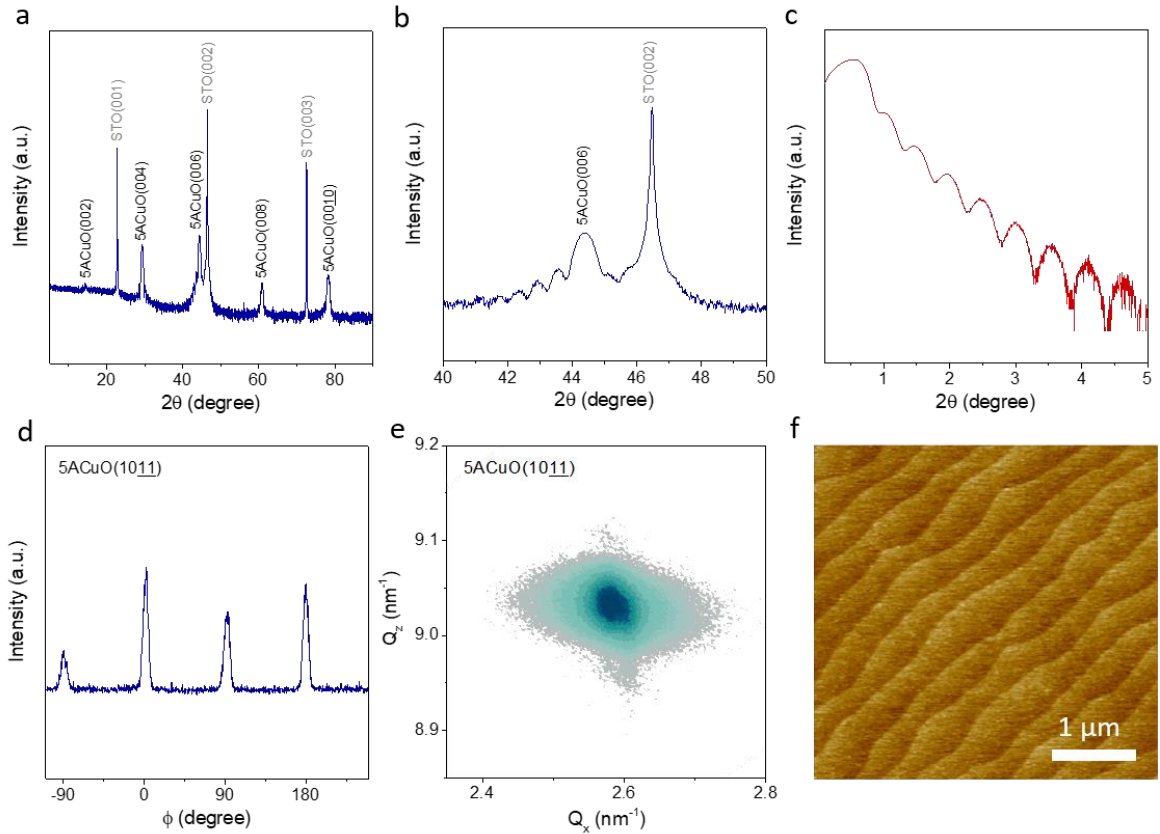

**Figure 1**. (a) $\theta-2\theta$ XRD full scan, (b) local scan near the STO(002) diffraction peak and (c) XRR pattern of an epitaxial 20-nm-thick 5ACuO film on STO(001). (d) ϕ-scan and (e) RSM scan near an asymmetric (10$\underline{1}\underline{1}$) diffraction pattern for the 5ACuO film. (f) Topography AFM image of the 5ACuO film showing atomically smooth surface.

**2.2 Local cation distribution**

The cation arrangement of a 5ACuO film grown on STO is examined by scanning transmission electron microscopy (STEM) in the high angle annular dark field (HAADF) mode. Consistent with the XRD results, the 5ACuO film presents a single-crystalline lattice, which establishes an abrupt interface with the STO substrate (Fig. 2a). Atomic-resolution STEM reveals that the 5ACuO film shares the same atomic stacking sequence as singly *A*-site populated $X_2CuO_4$ (Fig. 2b), which consists of one $CuO_2$ plane alternatively sandwiched between two neighboring *X*O planes (Fig. 2c). The local cation distribution is further probed with atomic-resolution energy-dispersive X-ray spectroscopy (EDS) mapping. The element-specific EDS maps in Fig. 2d show clean separation of the $CuO_2$ plane and the five *A*-site cations in the neighboring blocking plane. Meanwhile, no preferred ordering is observed among *A*-site cations, suggesting a homogeneous and random cation distribution in the blocking layer.

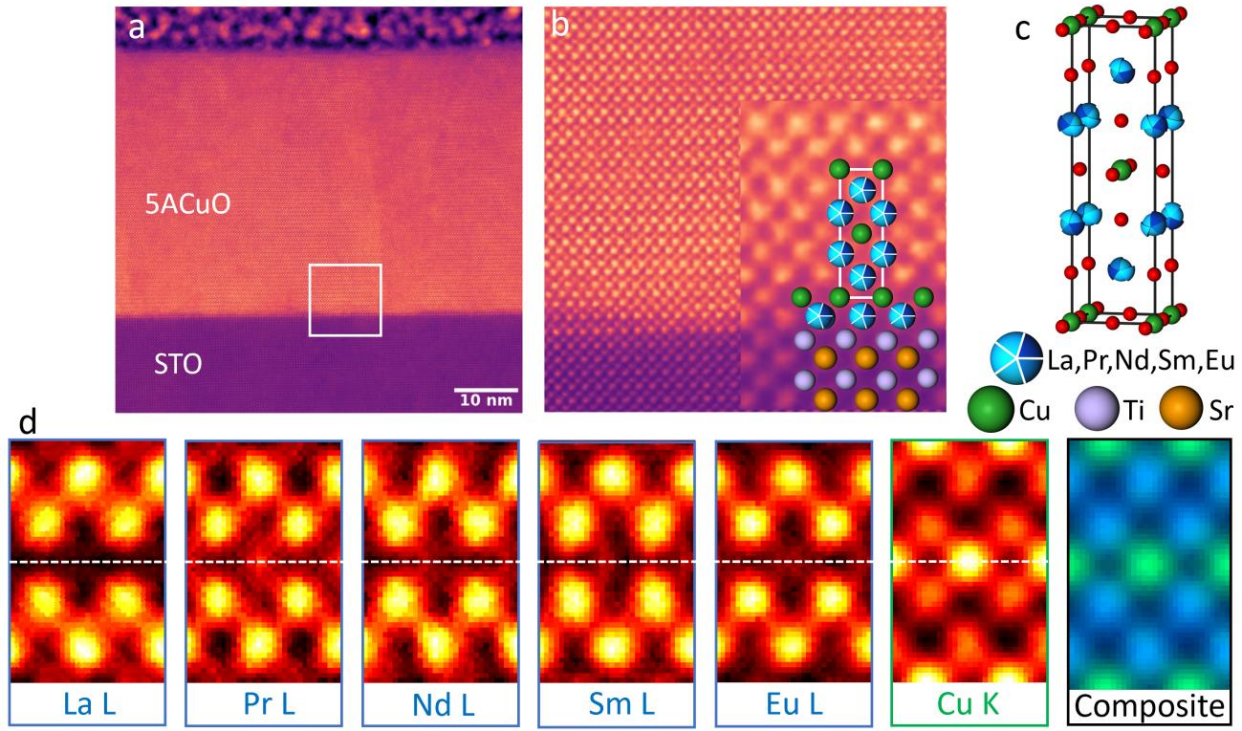

**Figure 2**. (a) Cross-sectional HAADF-STEM image of a 5ACuO film on STO(001) for the microstructure overview. (b) Atomic-resolution HAADF-STEM image demonstrating a single-crystalline film lattice. The inset shows an overlaid structural model of the film-substrate region. (c) Crystallographic model of the 5ACuO lattice. (d) Element-specific EDS maps of La-$L$, Pr-$L$, Nd-$L$, Sm-$L$, Eu-$L$, Cu-$K$ and the last color mix image demonstrating uniform mixing of $A$-site cations and clear alignment with the Cu plane.

**2.3 Electronic structure and chemical characterization**

Singly lanthanide $A$-site populated $A_2CuO_4$ can stabilize as three major tetragonal structures ($T$, $T^*$, $T'$) that feature distinct $CuO_x$ polymorphs.[32,33] The $T$-type structure is often associated with $La_2CuO_4$ and consists of octahedral $CuO_6$ units with six oxygens around the Cu site; through doping or annealing, the $T^*$-type structure can be observed in which one apical oxygen is disconnected which leaves five oxygens around the Cu site; the $T'$-type structure is often associated with $Nd_2CuO_4$ and consists of square-planar $CuO_4$ building blocks with four in-plane oxygens and no apical oxygen connections. The subtle change of local lattice deformation impacts orbital hybridization and electron transitions at the Cu site, which can be probed by X-ray near-edge absorption structure (XANES). The XANES spectrum near the Cu $K$-edge identifies a primary $T'$-type structure for 5ACuO films (Fig. 3a). Representative Cu $K$-edge XANES spectra of three types of cuprate structures from previous reports are also plotted as a reference.[32] These structures present distinct absorption features before and after the absorption edges, which correlate to $1s$-$4p\pi$ and $1s$-$4p\sigma$ transitions, respectively.[33] Compared to the octahedrally coordinated Cu sites, the square-planar Cu sites in the $T'$-type structure feature a more prominent pre-edge peak A due to the presence of ligand-to-metal charge transfer.[33] This is clearly seen in 5ACuO films which signals that the $T'$-type structure appears to dominate in films grown on STO.

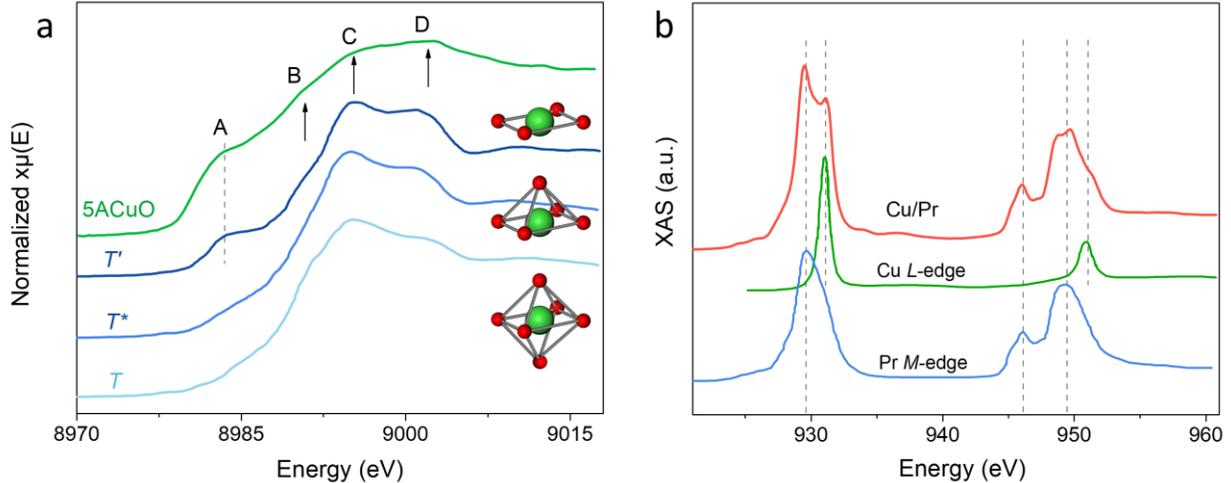

**Figure 3**. X-ray spectroscopy of a 5ACuO film on STO (001). (a) Cu $K$-edge XANES spectrum of the film compared to reference spectra [32] of three representative cuprate structures, $T$, $T^*$, $T'$ as shown in the inset. (b) XAS spectra around the Cu $L$-edge and Pr $M$-edge of the 5ACuO film show overlaps in energy range consistent with the superposition of reference spectra [34,35] of individual Cu $L$-edge and Pr $M$-edges resulting from the presence of $Pr^{3+}$ and $Cu^{2+}$.

Element-specific XAS in the soft X-ray range is also used to investigate the oxidation state of individual cations in the 5ACuO films. Fig. 3b presents the total electron yield (TEY) XAS spectra of the 5ACuO film on STO. Bulk-sensitive total florescence yield (TFY) was also collected, and while this data suffered from expected self-absorption effects that limit quantitative comparisons, a comparison of the $M_4$-edge features from the TEY and TFY spectra show that there is no significant qualitative feature differences, which indicates excellent depth-dependent homogeneity. Analyses of the Nd-*M*, Sm-*M*, Eu-*M*, and Pr-*M* -edge features (Fig. S4, supporting information) indicate all are in the 3$^+$ state. [34,35] The La-*M* and Cu-*L* -edge XAS features are indistinguishable from the well-studied $Nd_2CuO_4$ compound with the *T'*-type structure. As the Cu *L*-edge and Pr *M*-edges overlap significantly, the spectral features for the Cu and Pr are identified using specific reference spectra as shown in Fig. 3b. This allows for a satisfactory description of the Pr and Cu spectra as $Pr^{3+}$ ($Pr_2O_3$) and $Cu^{2+}$ (CuO) in the d$^9$ configuration.[34,35] No significant satellite peaks are detected in the high-energy shoulder on the Cu spectra, which would be present in a hole doped cuprate.[36]

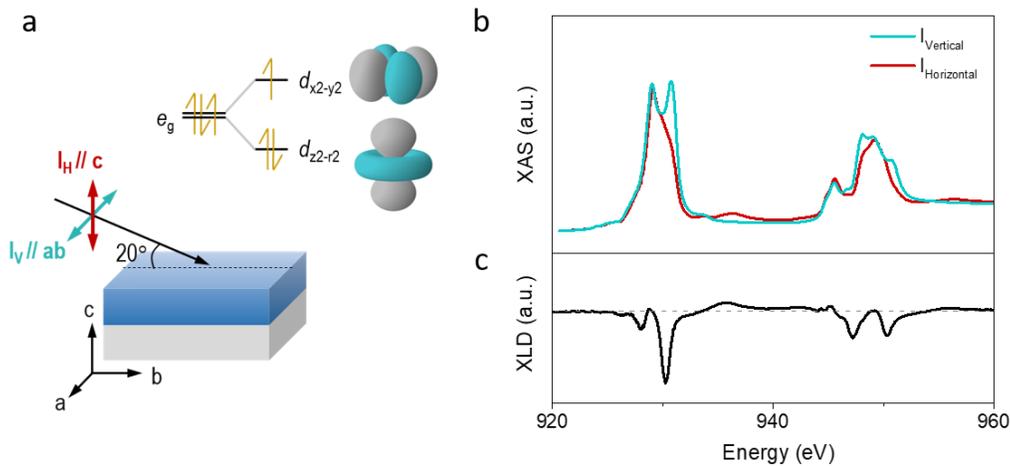

**Figure 4**. X-ray linear dichroism of a 5ACuO film on STO. (a)Schematic of the XLD measurement with the incident polarized X-ray beam at vertical and horizontal geometries and a schematic of the electronic configuration of Cu. (b)Vertical and horizontal polarization XAS spectra near the Cu *L*-edge and the Pr *M*-edge and (c) resulting XLD.

To further understand the Cu local electronic environment, we collected the x-ray linear dichroism (XLD) by measuring the absorption with horizontal ($I_H$) and vertical ($I_V$) polarizations. As shown in Fig. 4a, based on the grazing geometry of the experiment, $I_H$ and $I_V$ arise from the unoccupied Cu $d_{z2-r2}$ and $d_{x2-y2}$ states, respectively. Fig. 4b shows clear Cu orbital polarization with $I_H < I_V$. The linear dichroism (XLD = $I_H$ - $I_V$) in Fig. 4c demonstrates Cu orbital polarization typical of the $d$-electron occupancy favoring the $d_{z2-r2}$ orbitals [36,37,38] A direct comparison with the XLD of other oxide superlattices hosting the planar cuprate with the $3d^9$ configuration[40] is consistent with the observed XLD features, which further supports the preference of the $d_{z2-r2}$ orbital character that dominates in the *T'* structure. The details of the spectral features from soft XAS are not perfectly sensitive to local structure, however when comparing each of the characterizations above the dominant signature in all cases is consistent with the 5ACuO films grown on STO being of the 2D-like RP *T'* phase.

**2.4 Tailoring phase evolution with epitaxial templating**

The level of configurational complexity present in the 5ACuO material suggests that many other possible crystal phases could be formed from this combination of cations. The stabilization of such a complex crystal as the Ruddlseden-Popper structure is somewhat surprising considering the many possible and far simpler parent phases. An explanation for this is likely related to previous studies based on entropy stabilized oxide bulk ceramics, which suggest that entropy stabilization can act to dominate phase stabilization.[22,24] To begin to understand how structural scaffolding imparted by the substrate might impact crystal phase formation, a range of substrates are used during synthesis under otherwise identical growth conditions. At very high lattice mismatch, where the substrate lattice parameter ($a_{sub}$) increases to 4.211 Å on a MgO substrate, the 5ACuO film becomes amorphous; while this is likely due to the significant lattice mismatch, it is worth noting that MgO substrates are notorious for extrinsic surface roughness which might also be a contributing factor. For near matched substrates such as STO and DyScO$_3$ (DSO), the RP phase is stabilized with excellent uniformity and crystallinity (Fig. 5a-c). The $a_{sub}$ ranging between 3.90 - 3.95 Å appears to represent an optimum regime for RP phase synthesis with this

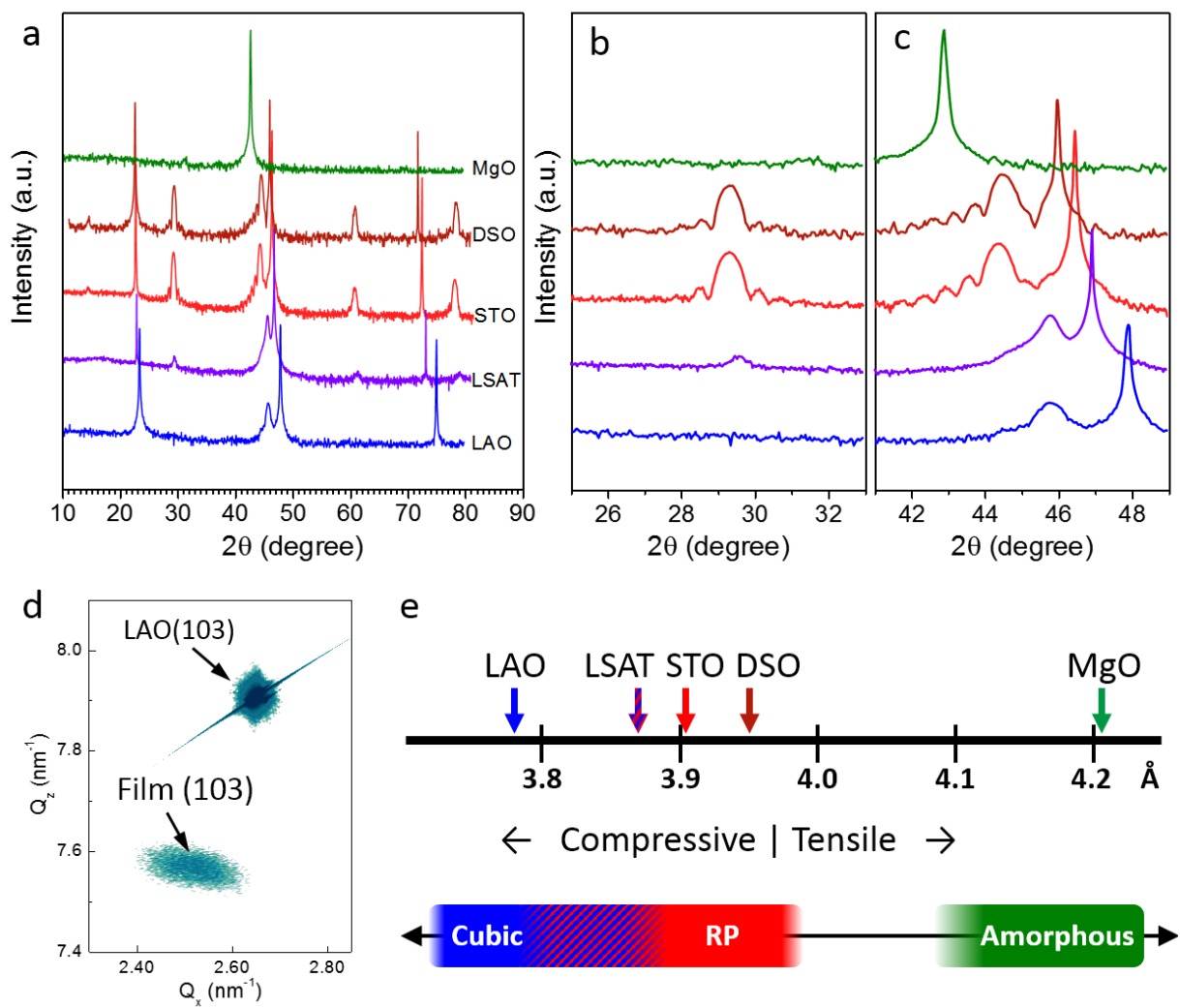

**Figure 5**. Comparison of films grown on a range of substrates at otherwise identical conditions. (a-c) $\theta-2\theta$ XRD scans of 5ACuO films grown on substrates with different lattice parameters. (d) RSM scan near LAO (103) for a film grown on LAO demonstrating a cubic crystal structure with lower symmetry than a simple rocksalt phase. (e) Schematic of phase evolution for the 5ACuO films tuned by the underlying substrate lattice parameter.

stoichiometry. Increasing compressive strain by growing on LSAT, greatly reduces the RP phase peak intensities and leads to a second phase formation. Intriguingly, as $a_{sub}$ is further reduced to 3.787 Å as in the case of LaAlO$_3$ (LAO), the 5ACuO film exhibits a single diffraction peak near the LAO (002) peak. The φ-scan of this film peak reveals a characteristic four-fold symmetry, suggesting the cube-on-cube epitaxial relationship between the 5ACuO film and LAO (Fig. S5,

Supporting Information). Asymmetric XRD scans further point toward a cubic crystal structure for this emerging phase, which exhibits a distinct film peak near the asymmetric LAO (103) peak in the RSM scan (Fig. 5d). From these structural measurements, we can state that the phase is not consistent with a rocksalt phase due to symmetry considerations.[41] Nor is this new phase a change in the $n$ value of the RP phase. The phase is a rather large cubic phase and is consistent with a spinel structure of a form $Cu(5A)_2O_4$. While further work is needed to fully map the microstructure of this cubic phase and its growth mechanism, the trends observed reveal that substrate selection plays a significant role in setting the film's crystal phase during synthesis (Fig. 5e). Also of note is the observation that increasing a coherently grown film beyond its relaxation thickness does not result in a degradation in film quality or a change in phase—single phase, uniformly orientated crystallinity is maintained even after the film begins to relax from its starting point. The implications of this may be that the stoichiometry of the target is preserved during synthesis but crystal phase can be selected during the initial crystallization by selecting a suitable substrate scaffold. The exact role of each of the mechanisms driving this initial phase formation remain an open question.

## 3. Conclusion

This work demonstrates the synthesis of $(La_{0.2}Pr_{0.2}Nd_{0.2}Sm_{0.2}Eu_{0.2})_2CuO_4$ Ruddlesden-Popper films. In-depth structural and chemical characterizations show that despite the complexity of this system, single crystalline films with atomically abrupt film-substrate interfaces and atomically smooth film surfaces can be synthesized using pulsed laser deposition. The RP-phase films have uniformly mixed lanthanide distribution on the $A$-site blocking layer while maintaining a homogenous copper oxide plane. XANES and XLD identify signatures of the characteristic $T$'-type RP structure, which is defined by its 2D square-planar $CuO_4$ planes separated by configurationally complex lanthanide oxide blocking layers. By synthesizing films on a range of substrates under identical growth conditions, it is found that heteroepitaxy can be used to control the crystal phase. Whereas weak tensile and near lattice matched conditions lead to the RP phase, the application of compressive strain changes the preferred phase toward a cubic structure which is consistent with a $CuX_2O_4$ spinel. Manipulation of configurational complexity provides a new route to understanding electron-phonon couplings and novel correlated *f*-electron physics in 2D layered RP structures. The added capability of selecting between RP and 3D cubic crystal structures will have a broad impact on functionalities dominated by crystal structure and disorder effects, such as magnetoresistance, unconventional superconductivity, ferroelectricity, catalysis, and ion conductivity.

**Acknowledgements**

Experiment design, sample synthesis, and structural characterization were supported by the US Department of Energy (DOE), Office of Basic Energy Sciences (BES), Materials Sciences and Engineering Division. Part of the STEM characterization was conducted through user proposal at the Center for Nanophase Materials Sciences, which is a US DOE, Office of Science User Facility. This research used resources of the Center for Functional Nanomaterials and the Inner Shell Spectroscopy 8-ID beamline of the National Synchrotron Light Source II, which are U.S. DOE Office of Science User Facilities at Brookhaven National Laboratory under Contract No. DE-SC0012704. Sandia National Laboratories is a multi-program laboratory managed and operated by National Technology and Engineering Solutions of Sandia, LLC., a wholly owned subsidiary of Honeywell International, Inc., for the U.S. Department of Energy's National Nuclear Security Administration under contract DE-NA0003525. This paper describes objective technical results and analysis. Any subjective views or opinions that might be expressed in the paper do not necessarily represent the views of the U.S. Department of Energy or the United States Government.

**Conflict of Interest**: The authors declare no conflict of interest.

**Experimental**

**Thin film growth.** 5ACuO thin films with various thicknesses were grown on STO(001) single-crystal substrates using PLD. A KrF excimer laser (λ= 248 nm) was used with a frequency of 5 Hz and a laser fluence of 0.7 J/cm$^2$. The growth optimization was performed, and optimal deposition conditions were found at a growth temperature of 790 ℃ and an oxygen pressure of 0.1 mTorr. After deposition, the films were cooling down in the same atmosphere to room temperature at a cooling rate of 20 ℃/min. The thin film growth rate was estimated to be 0.004 nm per pulse. The laser ablation target was prepared in the stoichiometry of (La$_{0.2}$Pr$_{0.2}$Nd$_{0.2}$Sm$_{0.2}$Eu$_{0.2}$)$_2$CuO$_4$ using a conventional solid-state reaction method.

**Structural and chemical characterization.** The crystal structure and epitaxial relations of thin films and the ceramic target were characterized by XRD using a high-resolution four-circle diffractometer with Cu-$K_{\alpha 1}$ radiation. Rietveld refinement was performed using GSAS-II for deriving the crystal structure of the ceramic target. The film surface morphology was characterized by atomic force microscope (AFM, Nanoscope III). The STEM characterization was performed in the HAADF mode using a Nion UltraSTEM 200 operated at 200 kV. The atomic-resolution EDS mappings was performed using FEI Titan G2 800-200 STEM equipped with a Cs probe corrector and ChemiSTEM technology. The cross-sectional TEM specimen was fabricated using a focused ion beam lift-out technique (FEI Helios).

**X-ray absorption characterization.** XANES measurements were performed in a fluorescence mode at room temperature at beamline 8-ID inner shell spectroscopy (ISS) of National Synchrotron Light Source II, Brookhaven National Laboratory. Soft XAS measurements were performed at beamline 4-ID-C of Advanced Photon Source, Argonne National Laboratory. The XAS spectra were collected with bulk-sensitive fluorescence yield and surface-sensitive electron yield modes at 10 K. XLD measurements were performed with linearly polarized x-rays. The X-ray polarization vector was parallel to the in-plane or out-of-plane direction of the thin film sample, respectively.

# Figures

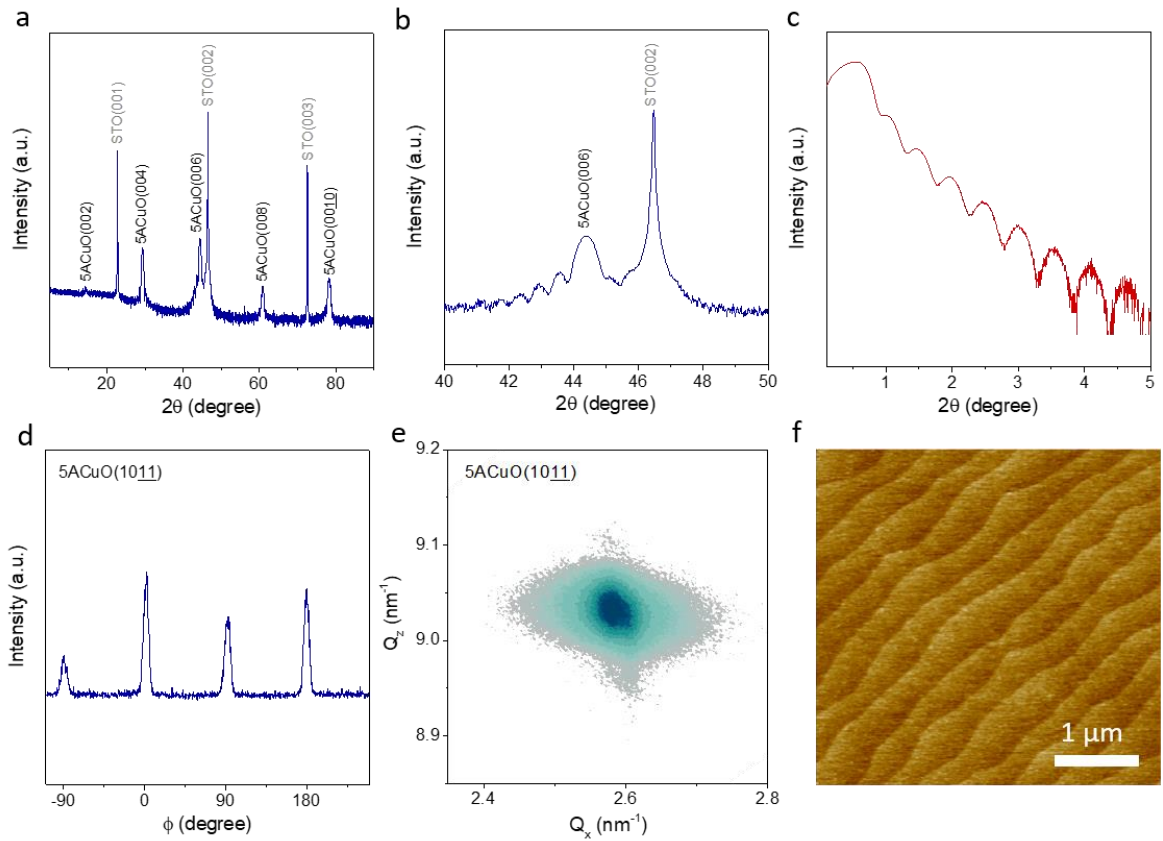

**Figure 1**. (a) $\theta-2\theta$ XRD full scan, (b) local scan near the STO(002) diffraction peak and (c) XRR pattern of an epitaxial 20-nm-thick 5ACuO film on STO(001). (d) $\phi$-scan and (e) RSM scan near an asymmetric (10$\bar{1}\bar{1}$) diffraction pattern for the 5ACuO film. (f) Topography AFM image of the 5ACuO film showing atomically smooth surface.

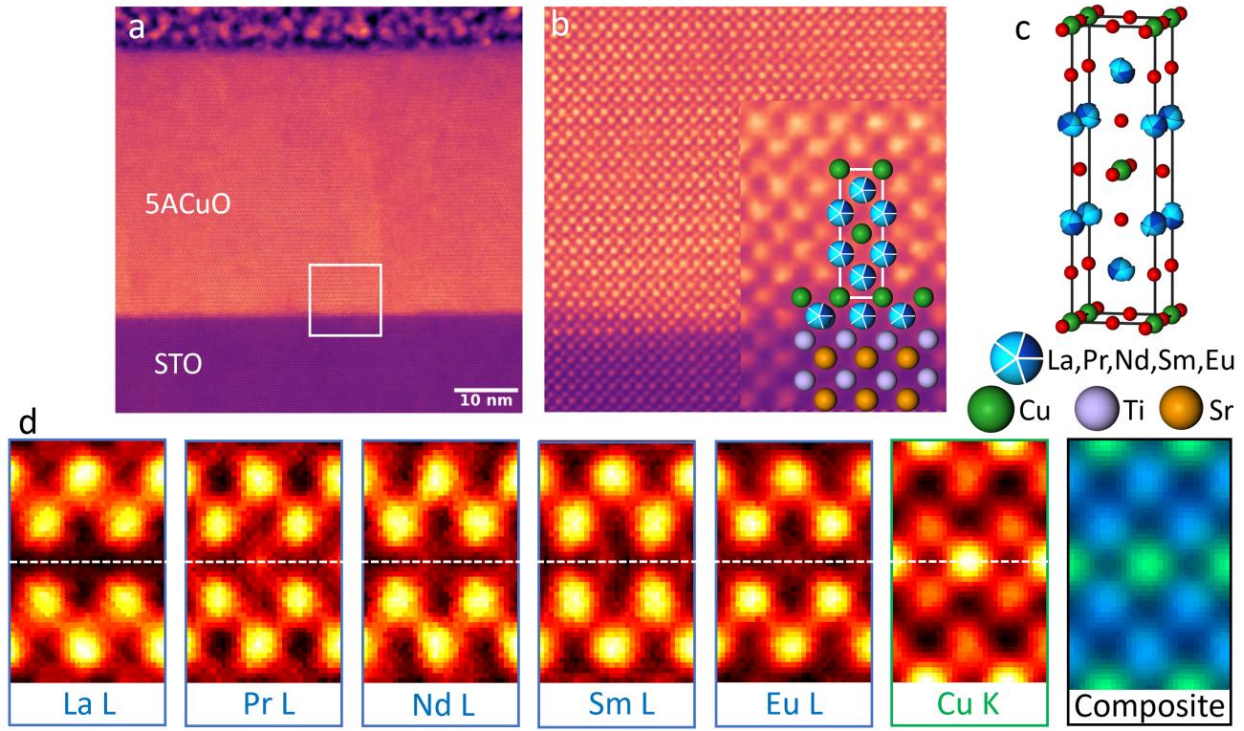

**Figure 2**. (a) Cross-sectional HAADF-STEM image of a 5ACuO film on STO(001) for the microstructure overview. (b) Atomic-resolution HAADF-STEM image demonstrating a single-crystalline film lattice. The inset shows an overlaid structural model of the film-substrate region. (c) Crystallographic model of the 5ACuO lattice. (d) Element-specific EDS maps of La-*L*, Pr-*L*, Nd-*L*, Sm-*L*, Eu-*L*, Cu-*K* and the last color mix image demonstrating uniform mixing of *A*-site cations and clear alignment with the Cu plane.

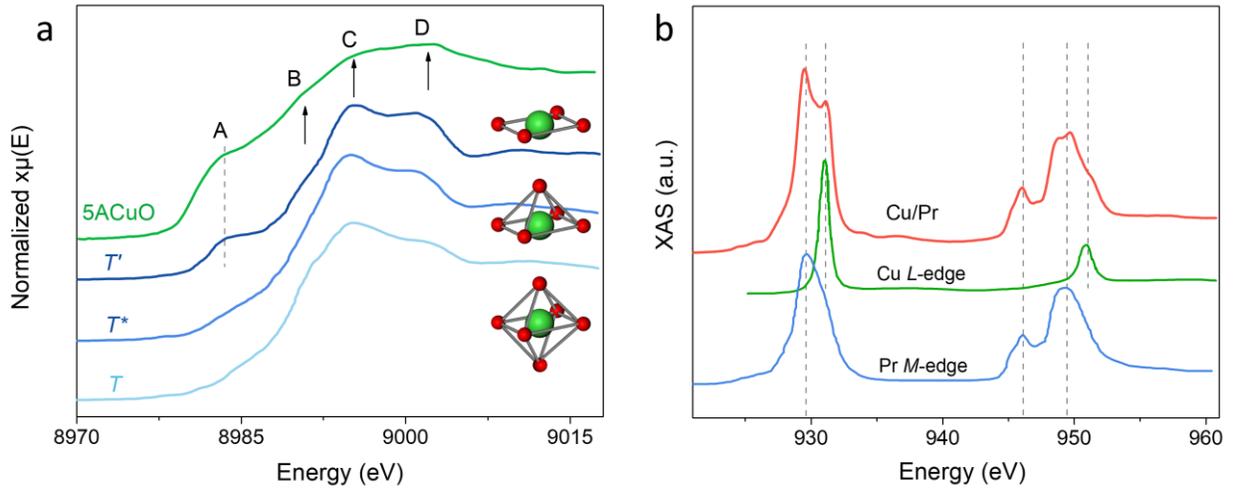

**Figure 3**. X-ray spectroscopy of a 5ACuO film on STO (001). (a) Cu *K*-edge XANES spectrum of the film compared to reference spectra [32] of three representative cuprate structures, *T*, $T^*$, $T'$ as shown in the inset. (b) XAS spectra around the Cu *L*-edge and Pr *M*-edge of the 5ACuO film show overlaps in energy range consistent with the superposition of reference spectra [34,35] of individual Cu *L*-edge and Pr *M*-edges resulting from the presence of $Pr^{3+}$ and $Cu^{2+}$.

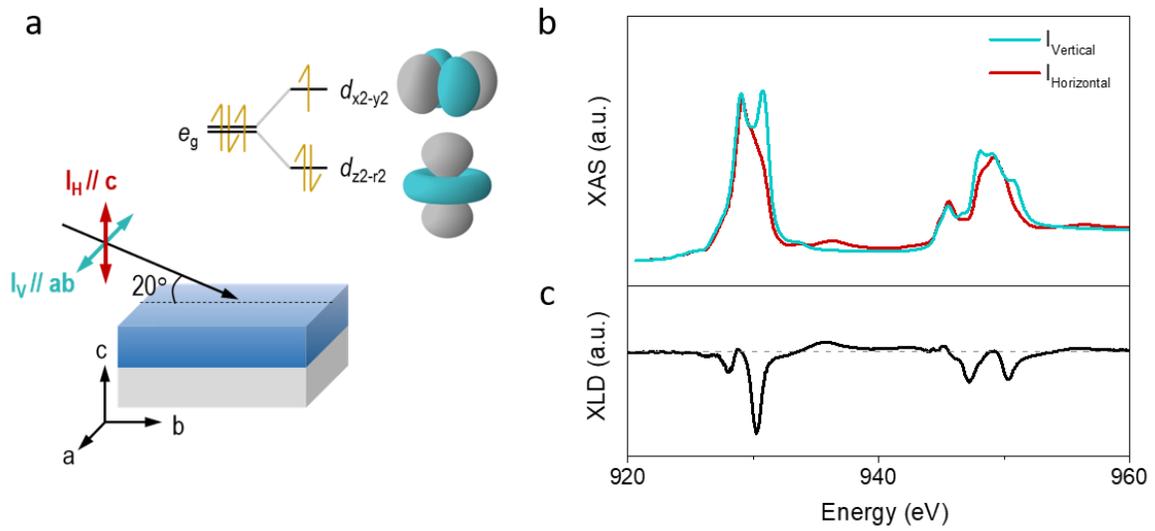

**Figure 4**. X-ray linear dichroism of a 5ACuO film on STO. (a)Schematic of the XLD measurement with the incident polarized X-ray beam at vertical and horizontal geometries and a schematic of the electronic configuration of Cu. (b)Vertical and horizontal polarization XAS spectra near the Cu *L*-edge and the Pr *M*-edge and (c) resulting XLD.

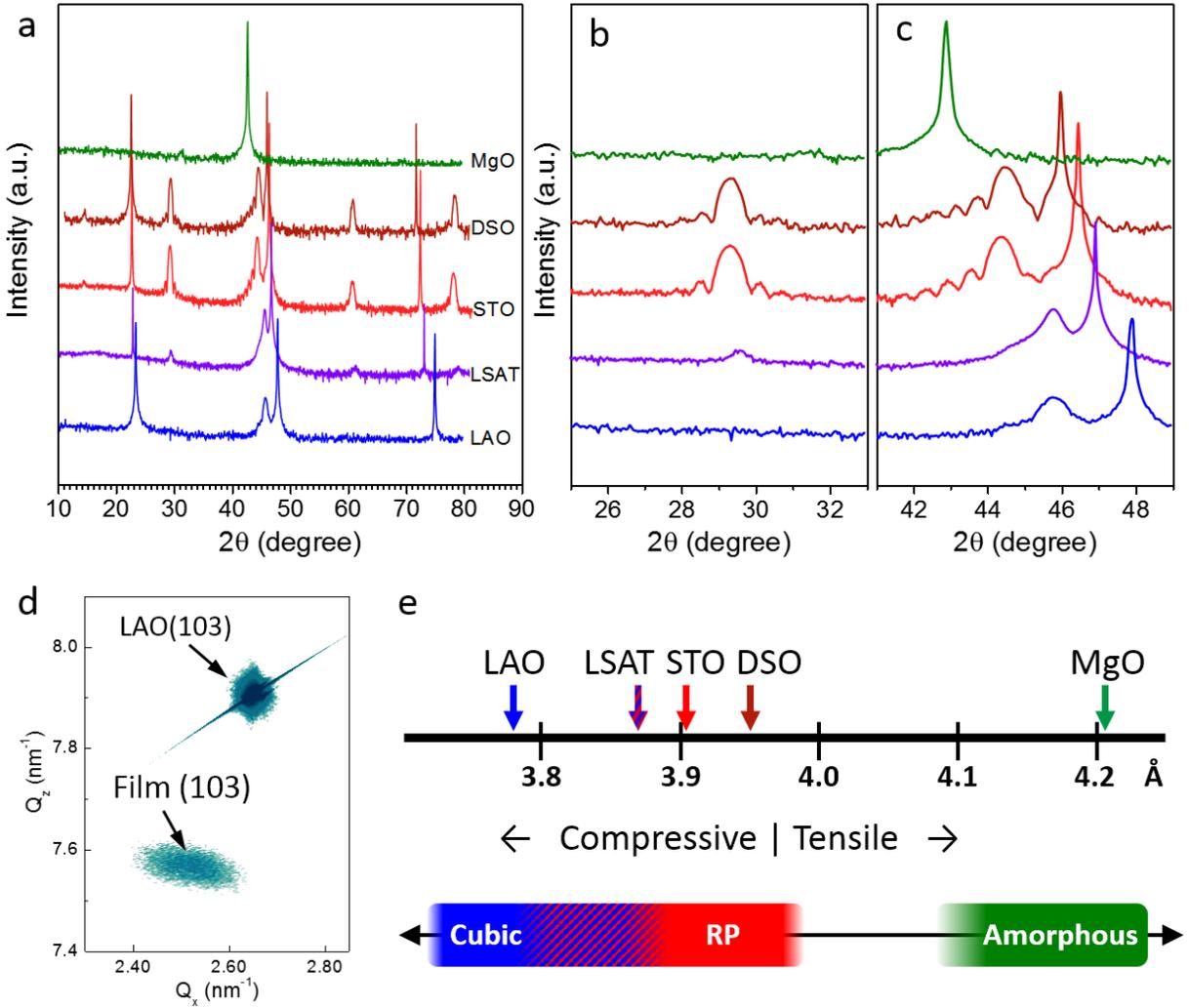

**Figure 5**. Comparison of films grown on a range of substrates at otherwise identical conditions. (a-c) $\theta-2\theta$ XRD scans of 5ACuO films grown on substrates with different lattice parameters. (d) RSM scan near LAO (103) for a film grown on LAO demonstrating a cubic crystal structure with lower symmetry than a simple rocksalt phase. (e) Schematic of phase evolution for the 5ACuO films tuned by the underlying substrate lattice parameter.